\documentclass[preprint,12pt]{elsarticle}

\usepackage{amssymb}
\usepackage{amsmath}

\begin{document}

\begin{frontmatter}



\title{Outbreak Patterns of the Novel Avian Influenza (H7N9)}


\author{Ya-Nan Pan}
\author{Jing-Jing Lou}
\author{Xiao-Pu Han} \ead{xp@hznu.edu.cn}

\address{Alibaba Research Center for Complexity Sciences, Hangzhou Normal University, Hangzhou 311121, China}

\begin{abstract}
The attack of novel avian influenza (H7N9) in east China caused a serious health crisis and public panic.
In this paper, we empirically analyze the onset patterns of human cases of the novel avian influenza and observe several spatial and temporal properties that are similar to other infective diseases. More deeply, using the empirical analysis and modeling studies, we find that the spatio-temporal network that connects the cities with human cases along the order of outbreak timing emerges two-section-power-law edge-length distribution, indicating the picture that several islands with higher and heterogeneous risk straggle in east China.
The proposed method is applicable to the analysis on the spreading situation in early stage of disease outbreak using quite limited dataset.

\end{abstract}


\end{frontmatter}



\section{Introduction}
\label{}

In the spring of 2013, a novel avian influenza, H7N9 virus, has broken out \cite{1,2,3,4,5,6,7}. This new type of influenza has affected at least 132 people (the total reported human cases until June 30, 2013) and resulted in a huge social panic and billions dollars economic losses, even though there is no evidence of human-to-human transmission has been found so far \cite{8,9}. And also, several virological researches indicated that H7N9 virus has potential to be a novel disease with strong human-to-human infection \cite{9,10}.

Comparing with other avian influenzas without the capability of sustained human-to-human transmission (e.g. H5N1), a notable difference between them is that the birds' symptom of H7N9 usually is inapparent \cite{9}, even though the spreading of H7N9 virus in birds and poultry is quite fast. Some typical warning signals for the outbreak of avian influenza, such as the death of large number of wild and domestic birds, can not be observed in the process of the H7N9's spreading. It causes many serious troubles in the monitor of the epidemic situation and the estimation of public health risk on H7N9, due to the fact that usually the spreading within birds and poultry is not visible until the outbreak of human cases. People have tried several measures, e.g. the wider virus test for wild birds and poultry in order to detect the spreading situation. However, these measures usually are costly to be taken.

Due to the absence of the capability of sustained human-to-human transmission, human H7N9 cases generally are occurred dispersedly in the spreading background of bird cases, and the region with higher epidemic level on bird cases usually has a higher risk to occur the bird-to-human infection. Therefore, we may has a chance to find out some useful information which is related to outbreak patterns of human cases.

In this paper, based on the empirical analysis for the spatio-temporal patterns of the human cases of H7N9and the association with the comparison of modeling results, we find that, the spatio-temporal network that connects each of two nearest human cases along the order of outbreak timing can efficiently reflect the pattern of the macro spreading situation.

\section{Dataset and patterns}

We collect the spatio-temporal information of all the reported human cases shown in the real-time reports of Chinese Center for Disease Control and Prevention (http://www.chinacdc.cn/jkzt/crb/rgrgzbxqlg\_5295/rgrqlgyp/) during the term from Feb. 19, 2013 (the onset date of the first human case), to April 30, 2013.
The total time-length of our recording is 70 days, which covers the term from the first onset of human case to the almost end of intensive outbreaks.
In this term, the total of recorded human cases and cities with human cases respectively are 128 and 36. In May and June of 2013, four human cases are reported, however, not included in the dateset, because they occurred sporadically and out of the main period of disease outbreaks.
For each case, the dataset records the location (the living city of each human case) and the date of symptom onset of the case.

\begin{figure}
\center
  \includegraphics[width=13.5cm]{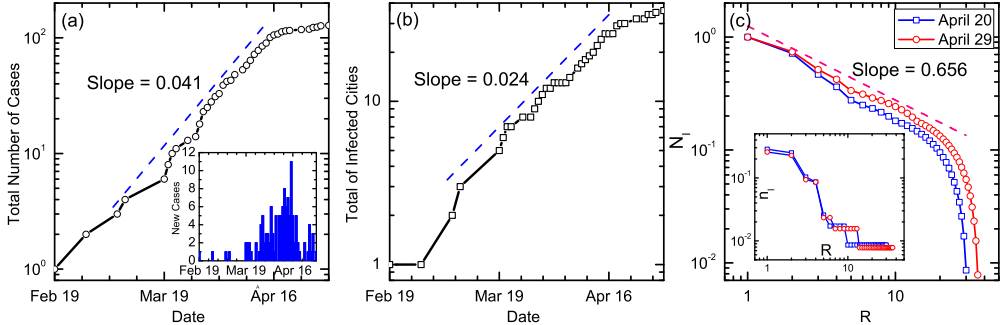}\\
  \caption{(Color online) (a) growth of the total of cases. (b) the growth of the total of infected cities. (c) the cumulative form of the distributions of normalized total cases in each city in Zipf's ranking plot for two dates, and the inset show its original form in Zipf's ranking plot, where the coordinate $n_I$ of the inset is the normalized total cases in each city, and the coordinate $N_I$ of the main panel is the cumulative number of $n_I$, namely, $N_I(R)= \sum_{r=R}^{R_{max}} n_I(r)$, and here $R_{max}$ is the total of infected cities.
  The dashed line in each panel shows the fitting exponent of each curve. }\label{growth}
\end{figure}

The total of cases grows exponentially before April 15, 2013 (Fig. \ref{growth}(a)), as well as the number of the infected cities (Fig. \ref{growth}(b)). After the peak in the middle of April (the inset of Fig. \ref{growth}(a)), the growth on both total of cases and infected cities tends to slow. Besides the patterns of growth, the spatial distributions of cases are also investigated. We find that the total of human cases in different cities matches power-law-like form in {\it Zipf's ranking plot} (see Fig. \ref{growth}(c) and its inset), indicating the heterogeneity in the outbreak of the new avian influenza for different regions: most of cases intensively outbreak in the region of a few cities, and most of cities only have one or two cases. The introduction of Zipf's ranking plot can be found in {\it Appendix A}.

Moreover, both of the exponential growth and scaling regional distributions are the typical features in the early outbreaks of many infectious diseases (e.g. SARS, the pandemic H1N1 in 2009 and the bird cases of avian influenza (H5N1))\cite{Han, WangL}, indicating the outbreak of human cases of H7N9 shows some similarities to infectious diseases, even though it has no the capability of sustained human-to-human transmission.
Since birds-to-human transmission is the major way for human cases, the spreading-like patterns would reflect that the spreading mode of H7N9 in birds and poultry is similar to the one of some human infective diseases. Due to the large difference between the organization of human society and avian ecosystem, the possibility that this similarity is resulted from human-activity-driven spreading (e.g. poultry trade and transportation) in birds/poultry can not be ruled out.

\section{Spatio-temporal Network of Cities with Human Cases}

For reasons of the difficulties in the wide detection of bird cases of H7N9 virus, the method that digs out useful information from the outbreak patterns of human cases would be of great value to the estimation of spreading situation of H7N9. Nevertheless, the sparsity of original data on human cases also rises the uncertainty and difficulty in the estimation by using the traditional methods (e.g. spatial density analysis). We therefore have to find a novel way to analyze the outbreak patterns of human cases. As the discussions below, the spatio-temporal network can be used here.

For a long time, network-based description is risen to be an important type of basic framework in the researches of epidemic spreading \cite{Moore,Newm,Pastor,Moreno,net1,net2,Gross}. The nodes of the network usually is to be the unit (e.g. individuals, cities, populations) in the spreading process, and the edges can express the paths that can transmit epidemic. Most of these network-based studies run the spreading process on networks, and few works use the inverse method that detect the real-world spreading patterns by using spatio-temporal networks, as we will do in this paper.
This type of inverse method is firstly proposed by Small et al \cite{Small} for the onset of bird cases of avian influenza (H5N1). The basic principle in the construction of the spatio-temporal networks is: to link each pair of outbreak locations that the distance shorter than a threshold relating to the time difference of the two outbreaks. Obviously, these links denote possible path in spreading, and the topology of the network reflects the basic infection relationships for different locations in the outbreak.

\begin{figure}
\center
  \includegraphics[width=7.9cm]{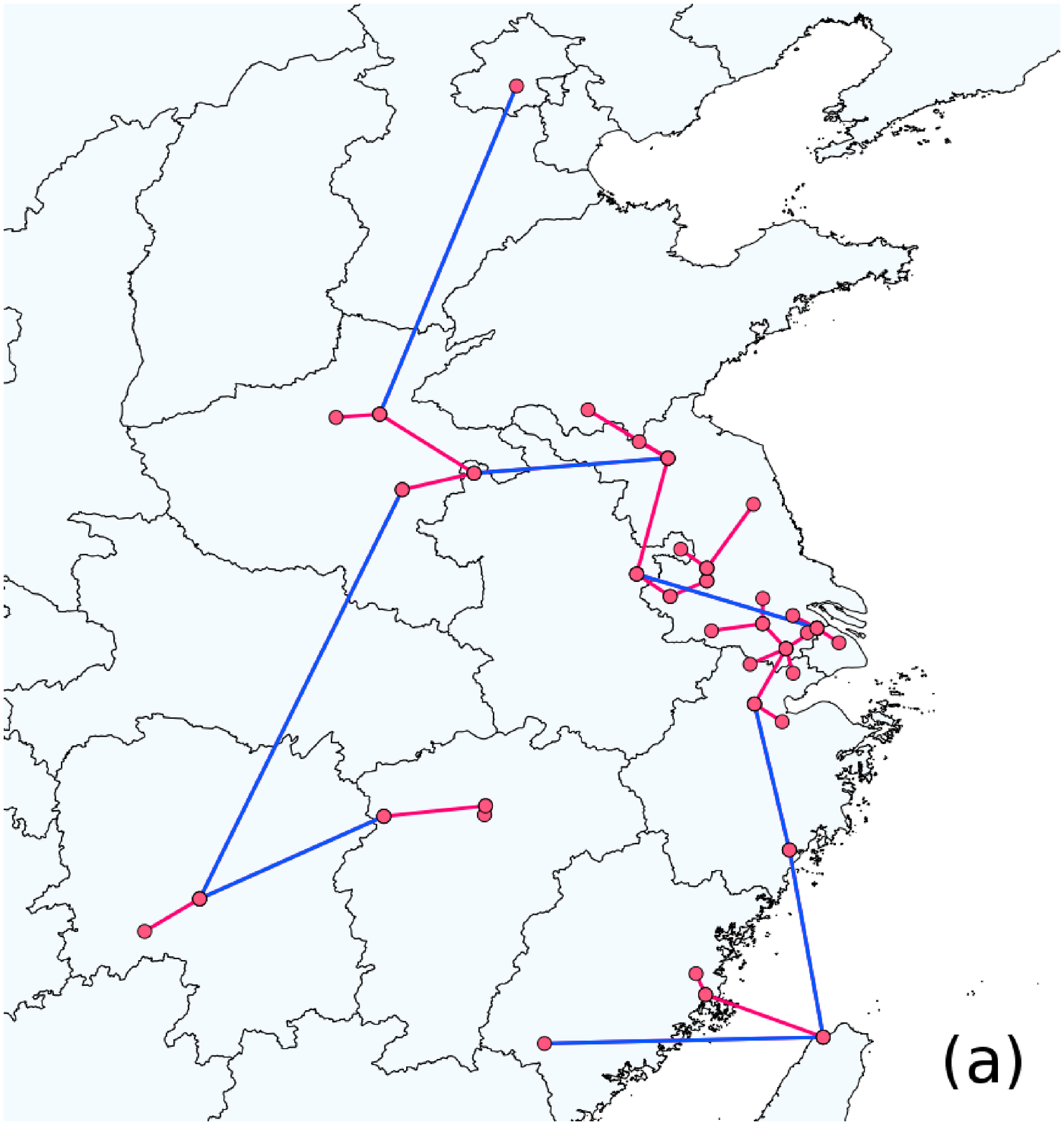}
  \includegraphics[width=5.6cm]{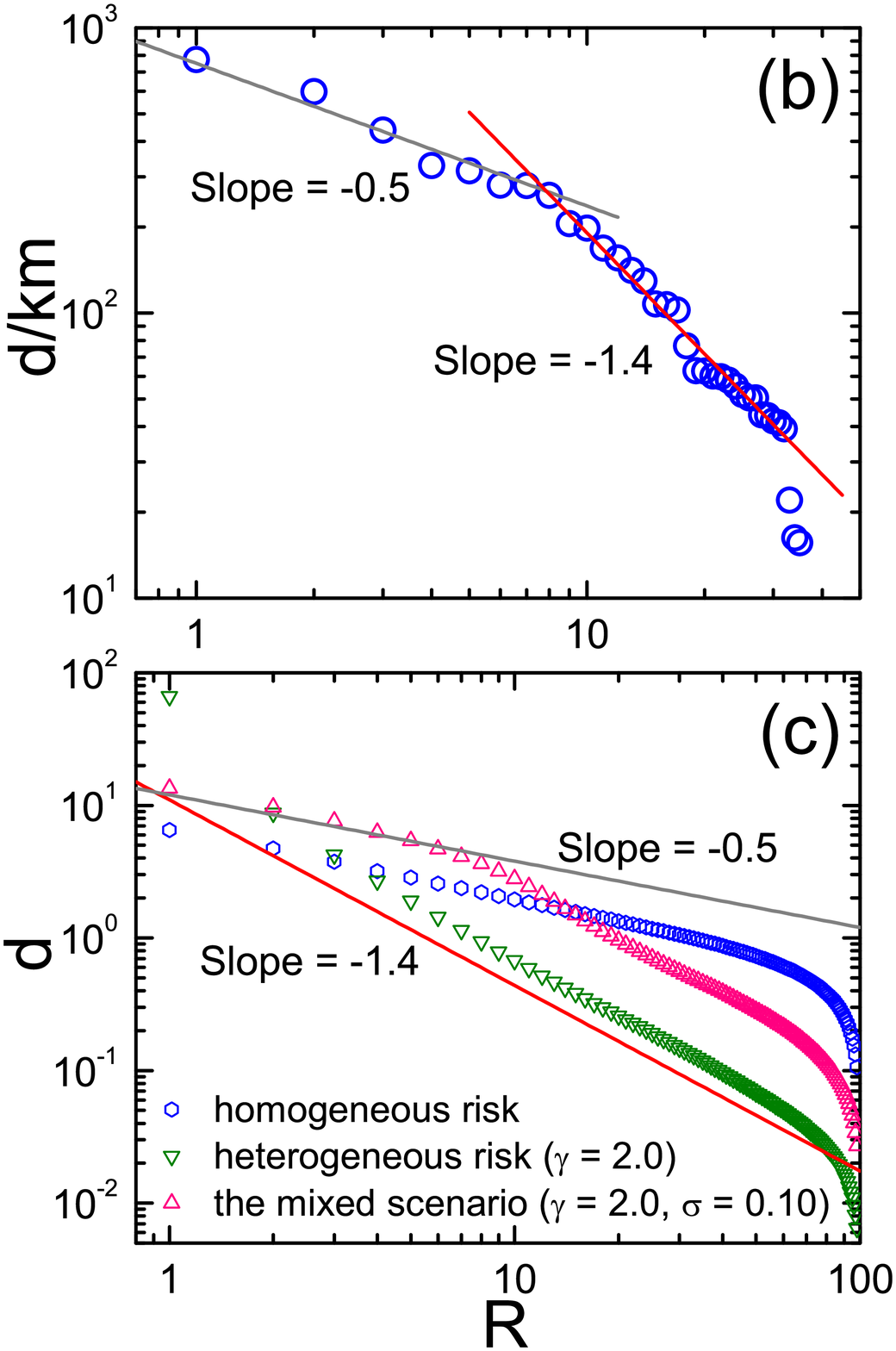}\\
  \caption{(Color online) (a) The CNENN of the novel avian influenza H7N9 in east China, where the gray and red lines respectively denote the edges with length $d \geq d_c$ and $d < d_c$ ($d_c = 259km$). (b) The two-section edge-length distribution of the real-world CNENN of H7N9 human cases in Zipf's ranking plot. (c) The edge-length distributions of modeling CNENN in Zipf's ranking plot for three situations on risk distribution: the homogenous case (blue cycles), the heterogeneous situation (green rectangles), and the mixed scenario (the pink rectangles). Each of them runs 100 time steps ($T = 100$), and all the results averaged over $10^3$ independent runs. To be comparable, the average edge length is fixed for different settings of the model. The gray and red lines respectively denote the power law with Zipf's exponent $0.5$ and $1.4$ in the empirical findings. }\label{spatial}
\end{figure}

Because the absence of direct transmission between human cases, our analysis focuses on the detection of higher-risk areas, and Small's method can not be used here directly. Generally speaking, if we connect each infected city to the nearest one, the shorter links will more frequently appear in the areas with higher onset risks (the density of infected cities is higher). Also considering the onset time of different cities, our method in the construction of the network is: to connect the new city with human cases to the nearest one from the cities that have broken out human cases before (so called {\it Connecting Nearest-Existing-Node Network}, and CNENN for short). The topology of CNENN therefore must be tree-like, as shown in Fig. \ref{spatial}(a). In a sense CNENN can be treated as the minimum spanning tree that connects the infected cities in space-time. Noticed that the links in CNENN do not denote the possible path in virus transmission, nevertheless, they reflect both the temporal relationships on onset time and the geographical density of infected cities.

And shown in Fig. \ref{spatial}(b), the edge lengths of CNENN of infected cities in the Zipf's ranking plot generally match the two-section power law:
\begin{equation} 
d \sim \left\{
        \begin{aligned}
            R^{-0.5}, (d \geq d_c) \\
            R^{-1.4}, (d < d_c)
        \end{aligned} \right.
\end{equation}
Where $R$ is the rank of edge lengths by descending order, and $d_c = 259km$. The edges in the two sections are shown in Fig. \ref{spatial}(a) respectively using blue and red lines, in which the short-range edges usually gather into several small fields that are connected by the long-range edges. The transition on the two section implies the factors driving the outbreaks in these small fields would be different to the one in the global level.

\begin{figure}
\center
  \includegraphics[width=13.5cm]{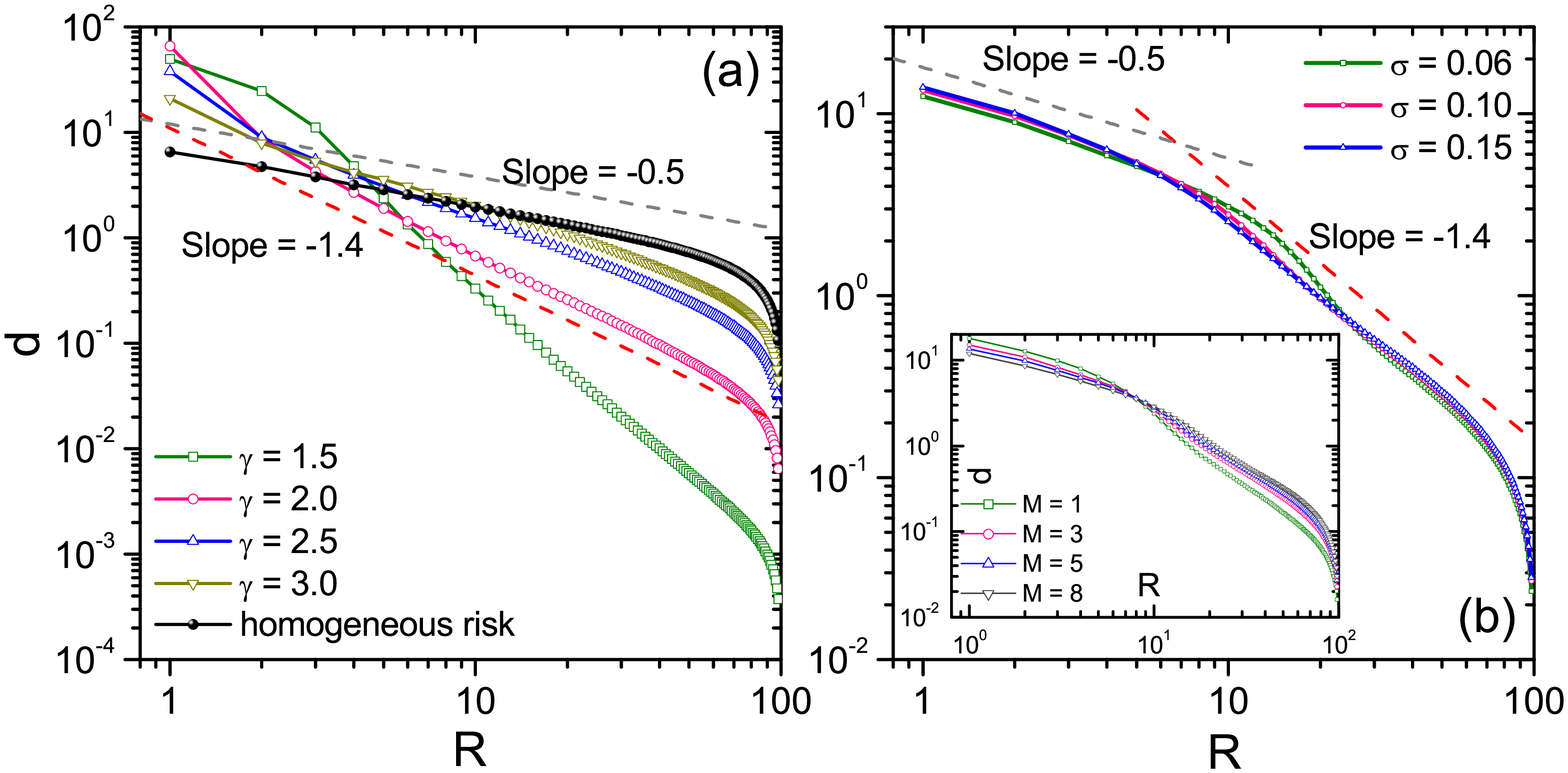}\\
  \caption{(Color online) (a) The edge-length distributions of modeling CNENN in Zipf's ranking plot for different heterogeneity ($\gamma$) and the situation with homogeneous risk. To be comparable, the average edge length is fixed for different settings of the model. (b) The edge-length distributions of CNENN of the mixed scenario for different $\sigma$ and different $M$ (the inset) when $\gamma = 2.0$. The gray and red dashed lines in the two panels respectively denote the power law with Zipf's exponent $0.5$ and $1.4$ in the empirical findings. All the results averaged over $10^3$ independent simulations, and each simulation runs 100 time steps ($T = 100$). }\label{model}
\end{figure}

To understand the origin of the two-section power law of CNENN, the empirical results are compared with the CNENN that is constructed by a simple model. In this model, each location (with coordinate $\vec{r}$, say) in a two-dimension area has a given risk value $\rho(\vec{r})$ to breakout cases. Initially ($t = 0$), none of locations have cases. At each time step ($t = 1, \cdots, T$), a new location ($\vec{r}$, say)that is randomly chosen with probability $\rho(\vec{r})$, breakouts cases. For $t > 1$, the new location creates an edge to connect the nearest one of the existed locations with cases. After $T$ time steps evolution, we have the CNENN of the modeling system that connects all the locations with cases.

The most simple situation of the model is that, all the locations have the same risk value $\rho$. In this situation, we must set a size $S \times S$ of the area. This situation is equivalent to the case that each location with cases is randomly placed on the area one by one. Numerical simulations and analytical results (see {\it Appendix B}) indicate that the edge-length distribution of the modeling CNENN in the homogeneous situation obeys the power law with Zipf's exponent $0.5$ in Zipf's ranking plot (Fig. \ref{spatial}(c)), which is same to the front section in Fig. \ref{spatial}(b), implying this section relates to the homogeneous nature in the risk of outbreak in the global level. That is to say, it reflects the background risk pervading in most regions of east China.

Considering the Zipf's law in regional distributions of human cases (Fig. \ref{growth}(c)), another situation of the model is assumed as that the spatial distribution of the risk $\rho$ is heterogeneous: $\rho(\vec{r}) \sim |r|^{-\gamma}$, where the parameter $\gamma$ presents the heterogeneity of risk (small $\gamma$ corresponds to large heterogeneity). As shown in Fig. \ref{model}(a), the edge-length distributions of the CNENN in the heterogeneous situation also obey power law in Zipf's plot, and the fitting Zipf's exponent $\beta$ reduces along the growth of $\gamma$ (reduce on heterogeneity of risk) and trends to $0.5$. The Zipf's exponent $1.4$ in empirical findings corresponds to the situation $\gamma \approx 2.0$ (see Fig. \ref{model}(a)), showing a quantified real-world risk heterogeneity.

A natural inference from above discussions is that, the origin of the observed two-section power law would relate to the co-effect of the above two situations. To insure it, the scenario mixing with both of the two situations is considered. In the mixed scenario, on the $S \times S$ area with homogeneous risk, we randomly place $M$ ``islands", and the radius of each island is $\sigma S$ ($0<\sigma<1$), where $\sigma$ is the relative radius of islands. The risk distribution inside each of the islands obeys the heterogeneous form $\rho(\vec{r}) \sim |\vec{r}-\vec{r_c}|^{-\gamma}$, where $\vec{r_c}$ is the position of the center of the island, and we set the risk $\rho(\vec{r})$ is higher than the one outside these islands. Here the overlap of several islands is allowed, and the risk in the overlapping area is the total of the corresponding risks in different islands. For example, for the position $\vec{r}$ in the overlapping area of $k$ islands, its risk is
$\rho(\vec{r}) \sim \sum_{i=1}^{k}|\vec{r}-\vec{r_{ci}}|^{-\gamma}$, where $\vec{r_{ci}}$ is the center of the i-th island. Simulations find that, the CNENN created by the mixed model shows two-section power-law-like edge-length distributions (Fig. \ref{model}(b)). The front section with Zipf's exponent $0.5$ extends along the decay on the relative radius of islands $\sigma$ and the growth on the number of islands $M$.
When $\gamma = 2.0$, the result can well fit the empirical findings (Fig. \ref{spatial}(c) and Fig. \ref{model}(b)).

Above all, the empirical findings and the modeling results show us the global outbreak picture on the infection risk of H7N9 virus: several islands with heterogeneous outbreak risk straggle in a large area with lower background risk in east China. These islands are the regions that have higher risk and need focus on in the control of the novel avian influenza (H7N9).
Due to the transition on the edge-length distribution of CNENN is $d_c = 259km$, $d_c$ would close to the upper limitation on the range of these islands.

\section{Discussions}

Using the dataset that only contains the information of 128 human cases and 36 cities with cases, the above empirical and modeling results uncover several basic properties in the outbreak of the novel influenza (H7N9).
Firstly, several basic spatial and temporal characteristics in the growth of human cases, including the exponential growth of total human cases and total cities with human cases, and the regional distributions of human cases, are mush similar to that of several infective diseases, reflecting the fast spread of the virus in birds and poultry, and implying the possible human-activity-driven mechanism in the spreading of H7N9 virus in birds and poultry. More importantly, on the discussion of the two-section power law of the edge-length distribution of CNENN of cities with human cases, the picture that several islands with higher and heterogeneous risk in the global spatial patterns of H7N9 are found out, even with the maximum range of the high-risk islands can be quantitatively estimated, indicating that the method using CNENN to detect the global spreading situation efficiently works in facing of the small-size dataset.

The core of CNENN method is the transformation from the density distribution of cities with human cases to the edge-length distribution of CNENN, which indeed introduces an averaged effect in the analysis: that is why the method is efficient here, although it still can not accurately locate the peak with the highest risk in each island. It also implies that the CNENN-like methods can be extended to be useful in the analysis of other systems using small-size datasets.
In summary, the CNENN method would be much useful in the detection and surveillance of spreading situation under the paucity of available data that we usually have to face in the early outbreak of many diseases.

\section{Acknowledgments}
\label{}
This work was funded by the National Natural Science Foundation of China (No. 11205040, 11105024, 11305043, 11275186, 10975126), and the Zhejiang Provincial Natural Science Foundation of China under the grant No. LY12A05003, and the research startup fund of Hangzhou Normal University.


\appendix

\section{Zipf's ranking plot and Zipf's law}

\emph{Zipf's ranking plot} is widely used in the statistical analysis of the small-size sample \cite{Zipf}, which can be obtained by first rearranging
the data by decreasing order and then plotting the value of each data point versus its rank.  Actually, Zipf's ranking plot shows the inverse form of the cumulative distribution of the data.

\emph{Zipf's law} describes a scaling relation in Zipf's ranking plot, $z(R)\sim R^{-\alpha}$, between the value of data point $z(R)$ and its rank $R$, where the exponent $\alpha$ usually is called ``Zipf's exponent".
As a signature of complex systems, the Zipf's law is widely observed in much social and natural systems \cite{Newman2005,Clauset2009}. It corresponds to a power-law probability density function $p(z)\sim z^{-\beta}$ with $\beta=1+\frac{1}{\alpha}$ \cite{Lv,Lv2,Gong}.

\section{Analysis of CNENN in the situation with homogeneous risk}

For the situation with homogeneous risk, an equivalent model is that, we randomly place a location with cases in the area at each time step.
Assuming there has been $m - 1$ locations appear cases in a square area with size $S \times S$, for the $m$-th location, it should choose the nearest one from the $m - 1$ existed locations to create a new edge (the length is $d$). So the probability density of the relative length $l = d/S$ of the new edge is equal to the probability that appears such situation: $m - 2$ existed locations are placed outside the ring centered by the new location with radius $d$, and one existed location is exactly right on the ring.
Ignoring the boundary effects of the square, we easily write out the normalized probability density of $l$ \cite{Han2},
\begin{equation}
\Phi(l) \approx (m-1)\times 2\pi l (1 - \pi l^2)^{m-2},
\end{equation}
where $(m -1)$ is the normalization factor of the distribution. From
\begin{equation}
\frac{\partial P(l)}{\partial l}|_{l=l_*} = 0,
\end{equation}
the maximum likelihood edge-length is
\begin{equation}
l_*=[(2m-3)\pi]^{-1/2}.
\end{equation}
For large $m$, $l_* \propto m^{-1/2}$. Due to the value $m$ denotes the current number of locations with cases, $l_*$ reduces along the growth of $m$, and thus $m$ approximately corresponds to the rank of edge-length. Accordingly, we have the edge-length distribution of CNENN in Zipf's ranking plot $d \sim R^{-1/2}$, as shown in Fig. \ref{spatial}(c), in agreement with the numerical simulations. Therefore the Zipf's exponent $0.5$ of CNENN corresponds to the situation with homogeneous risk.




\bibliographystyle{elsarticle-num}







\end{document}